\documentstyle[multicol,epsfig,aps]{revtex}
\def\be {\begin{equation}}
\def\ee {\end{equation}}
\def\ba {\begin{eqnarray}}
\def\ea {\end{eqnarray}}
\def\nn {\nonumber}
\def\a  {\alpha}
\def\b  {\beta}

\def\O  {\Omega}
\def\p  {\pi}

\def\s {\sigma}
\def\t  {\tau}
\def\la {\label}
\def\le {\left}
\def\ri {\right}
\def\pa {\partial}
\def\f {\frac}

\def\no {\noindent}
\def\bi {\begin{itemize}}
\def\ei {\end{itemize}}

\def\vs {\vspace}
\def\hs {\hspace}

\def\cals {{\cal S}}
\def\sbh  {\f{4S_{BH}}{\O_{d-2} R^{d-2}} }
\begin{document}
\draft
\title{Entropy Corrections for Schwarzschild and
Reissner-Nordstr\"om Black Holes}
\author{M. M. Akbar$^\flat$,~Saurya Das$^\sharp$
\footnote{Address from August 2003: Dept. of Physics, University 
of Lethbridge, 4401 University Drive, Lethbridge,
Alberta T1K 3M4, CANADA. EMail: saurya.das@uleth.ca  }}
\address{\hs{-8.6cm}${}_\flat$}
\address{Department of Applied Mathematics and Theoretical Physics,}
\address{Centre for Mathematical Sciences,}
\address{University of Cambridge,}
\address{Wilberforce Road,}
\address{Cambridge CB3 0WA, U.K.}
\address{EMail: M.M.Akbar@damtp.cam.ac.uk}
\address{\hs{-6.1cm}${}_\sharp$}
\address{Department of Mathematics and Statistics, }
\address{University of New Brunswick,}
\address{Fredericton, New Brunswick E3B 5A3, CANADA }
\address{EMail: saurya@math.unb.ca   }

\maketitle

\begin{abstract}
Schwarzschild black hole being thermodynamically unstable, corrections
to its entropy due to small thermal fluctuations cannot be
computed. However, a thermodynamically stable Schwarzschild solution
can be obtained within a cavity of any finite radius by immersing it
in an isothermal bath. For these boundary conditions, classically
there are either two black hole solutions or no solution. In the
former case, the larger mass solution has a positive specific heat and
hence is locally thermodynamically stable.  We find that the entropy
of this black hole, including first order fluctuation corrections is
given by: ${\cal S} = S_{BH} - \ln[\f{3}{R} (S_{BH}/4\p)^{1/2}
-2]^{-1} + (1/2) \ln(4\p)$, where $S_{BH}=A/4$ is its
Bekenstein-Hawking entropy and $R$ is the radius of the cavity. We
extend our results to four dimensional Reissner-Nordstr\"om black
holes, for which the corresponding expression is: ${\cal S} = S_{BH} -
\f{1}{2} \ln [ { (S_{BH}/\p R^2) ( {3S_{BH}}/{\p R^2} -
2\sqrt{{S_{BH}}/{\p R^2 -\a^2}} ) \le( \sqrt{{S_{BH}}/{\p R^2}} - \a^2
\ri) }/ {\le( {S_{BH}}/{\p R^2} -\a^2 \ri)^2 } ]^{-1} +(1/2)\ln(4\p).$
Finally, we generalise the stability analysis to Reissner-Nordstr\"om
black holes in arbitrary spacetime dimensions, and compute their
leading order entropy corrections. 
In contrast to previously studied examples, we find that 
the entropy corrections in these cases have a different character.

\end{abstract}

\begin{multicols} {2}

\section{Introduction}
It is well-known that the specific heat of a Schwarzschild black hole
is negative:
\be
C = \f{dM}{dT} = -8 \p M^2 < 0~. \label{specificheat}
\ee
Thus it is thermodynamically unstable, and corrections to thermodynamic
quantities of this black hole, e.g. entropy and temperature, are not
well-defined. This is seen when one considers for instance
the corrected entropy
of any thermodynamic system (including black holes) due to
small thermal fluctuations of the system, which is given by
\cite{dmb,saurya}:
\ba
{\cal S} &=& S_0 - \f{1}{2} \ln \le( <E^2> - <E>^2 \ri) + \cdots \nn \\
&=& S_0 - \f{1}{2} \ln \le( CT^2 \ri) + \dots~~~.
\la{master0}
\ea
When relation (\ref{master0}) is applied to
black holes, one substitutes $S_0 \rightarrow S_{BH}$,
the Bekenstein-Hawking
entropy of the black hole, and calculates $C$ for the particular
black hole under consideration.
Note that in using (\ref{master0}) in the context of black holes, one is 
simply assuming that the black hole behaves as an ordinary thermodynamic
system, following usual laws of thermodynamics. 
Then starting from a continuum partition function and performing an 
inverse Laplace transform gives the density of states, whose logarithm 
yields Eq.(\ref{master0}) above.
In addition, if one assumes
that the black hole (more precisely, its horizon) is built up of more
fundamental entities, such as quanta of area, or that observables such as
horizon area and mass are quantised in a certain way, 
then as shown in refs.\cite{gmed} and \cite{btd},
there could be additional terms in (\ref{master0}).   

It may also be noted that there could be other corrections to black hole
entropy due to a variety of other sources. For instance, quantum fluctuations
of matter fields in black hole backgrounds, as well as fluctuations of 
spacetime geometry itself in the canonical quantum gravity framework 
give rise to corrections which are also logarithmic in nature. Details of 
these calculations can be found in 
\cite{fursaev,solo1,other,nojiri,qg,obregon,carlip,jy,qg1,bss,gks,k,ms,medved,k2,medved2,gour1,cm,husain}.
Since our analysis is not tied to any specific model of quantum gravity, it 
remains valid as long as the black hole is large.
On the other hand, the other corrections may become important
as the size of the black hole approaches the Planck scale. 

One sees that the formula breaks down for $C<0$.
However, as was first shown by York \cite{york1}, a positive specific
heat solution can be obtained by considering Schwarzschild black holes
within a {\it finite} $S^2$ cavity immersed in an isothermal bath. In
fact for such canonical boundary conditions there are either two or no
solutions depending on whether the temperature of the bath is above or
below a minimum value. The larger mass of the solutions has a positive
heat and the other has a negative specific heat. When one takes
the cavity radius to infinity keeping the wall-temperature of the
cavity fixed (i.e., not changing the temperature of the bath) the
positive specific heat solution disappears and only the negative
specific solution survives giving the known black hole solution. This
holds for higher dimensions as well \cite{mm}. In this article, we
consider the positive specific heat solution and compute leading order
corrections to its entropy. As we will see the logarithmic correction
term due to small thermal fluctuation is of a different nature unlike
in several other cases studied before
\cite{fursaev,solo1,other,nojiri,qg,obregon,carlip,jy,qg1,bss,gks,k,ms,medved,k2,medved2,gour1,cm,husain}.
In this case, the mass and specific heat of the black hole are
functions of the temperature {\it and} cavity radius, and the entropy
correction is more interesting and non-trivial, as we will see
below. The case of charged black hole, i.e., Reissner-Nordstr\"om
black holes, within a cavity was first studied in \cite{york2} and we
will find correction for their entropy as well for thermal
fluctuation. See also \cite{louko} for related work on thermodynamics
of Schwarzschild and Reissner-Nordstr\"om-AdS black holes. 
This paper is arranged in the following way. In the next
section, we briefly review the thermodynamics of Schwarzschild black
holes within a cavity and compute the entropy correction. In section
(\ref{secrn}), we extend our results to Reissner-Nordstr\"om black
holes in arbitrary spacetime dimensions.  Finally, we end with a
summary of our results and some further directions in sections
(\ref{secdis}).  We use units in which $G=c=\hbar=k_B=1$.

\section{Schwarzschild Black Hole in a Box: Entropy corrections}
Let us first consider a Schwarzschild black hole of mass $M$. Wick
rotating the Lorentzian time $t \rightarrow i \t$ one obtains
%
\ba
ds^2 &=& \le( 1 - \f{2M}{r} \ri) d\t^2 \nn \\
&+& \le( 1 - \f{2M}{r} \ri)^{-1} dr^2 + r^2 d\O_2^2~, \label{Sch}
\ea
where $d\O_{2}^2$ is the metric on unit $S^{2}$. The metric (\ref{Sch}) is
singular at $r=2M$. The singularity can be removed by periodically
identifying $\t$ with a period $8\pi M$. Hence the resulting metric is
complete and regular for $2M\leq r < \infty$. The fixed point set of
the Killing vector $\partial/\partial \t$ is a regular bolt of the
metric \cite{GH}. The inverse of the periodicity of the $\t$
coordinate gives the temperature of the hole measured at infinity.

\subsection{\it Isothermal Cavity}
Next, consider a spherical $S^2$ cavity of radius $R$ concentric with
the black hole horizon. The black hole temperature measured at
infinity is shifted by $(g_{00})^{-\frac{1}{2}}$, as one moves towards
the black hole. The local temperature of the $S^2$ cavity is therefore
given by
%
\be
\b =  T^{-1}= 8\p M \sqrt{1 - \f{2M}{R} }~.
\la{b1}
\ee
Now we ask the reverse question. Given a spherical cavity for radius
$R$ held at a temperature $\b^{-1}$, what are the black hole
solutions?  In other words what are possible solutions for $M$? This
was first considered by York \cite{york1} for four
dimensions. Recently this has been discussed for higher dimensions in
\cite{mm}. For the purpose of exposition we deal with the case of four
dimensions here and discuss generalisation to arbitrary dimensions in
the next section. It is not difficult to see that there are in fact
two black hole solutions or no solutions by considering Eq.(\ref{b1})
which can be written in the following form :
\be M^3 - \f{1}{2} R M^2 + \f{R\b^2}{128 \p^2} = 0~.  \la{cube} \ee
This is a cubic equation in $M$ and always admits a negative root and
a pair of positive or complex roots depending on the coefficients. The
coefficients are determined by the boundary variables $R$ and $\b$.
Defining new variables $x \equiv 2M/R$ and $\s = \b/4\p R$,
Eq.(\ref{cube}) can be cast in the following simple form:
\be
x^3 - x^2 + \s^2 = 0~. \label{tri}
\ee
When positive solutions exist, larger and smaller solutions of $x$
give larger and smaller hole respectively. Note that the positive
roots take values from zero to one (see figure 1).  It can be shown that
two positive roots occurs if and only if \cite{york1}
\be
\s \leq \f{2}{\sqrt{27}} ~.
\la{ineq1}
\ee
%
%
Thus one requires the temperature $T$ of the bath to be
\begin{equation}
T \geq \frac{\sqrt{27}}{8\pi}\frac{1}{R}. \label{ineq2}
\end{equation}
At the equality of (\ref{ineq1}) or (\ref{ineq2}) the two roots are
degenerate.  The larger solution therefore takes value within the
closed interval $[\frac{2}{3},1]$.  For completeness we mention
here the explicit solutions of (\ref{tri}) here. The exact solutions are:
%
\ba
x_1 &=& \f{1}{3} \le[ 1 - 2\cos \le( \f{\a}{3} + \f{\p}{3}\ri) \ri]~,
\la{x1} \\
x_2 &=& \f{1}{3} \le[ 1 + 2\cos \le( \f{\a}{3} \ri) \ri]~,
\la{x2}
\ea
which translate in terms of black hole mass as:
\ba
M_1 &=& \f{R}{6} \le[ 1 - 2\cos \le( \f{\a}{3} + \f{\p}{3}\ri) \ri]~,
\la{m1} \\
M_2 &=& \f{R}{6} \le[ 1 + 2\cos \le( \f{\a}{3} \ri) \ri]~,
\la{m2}
\ea
where
\be
\cos\a = 1 - \f{27}{2} \s^2 ~.
\ee
Evidently at the equality of (\ref{ineq1}), $\a = \p$,
the solutions coincide, and
\be
x = \f{2}{3}~~\Leftrightarrow~~M = \f{R}{3}~.
\ee
By calculating the Euclidean action of the solution it was shown in
\cite{york1} that the entropy of the black hole is unaffected by the
presence of the box and is given by the usual relationship \be S_{BH}
= 4\p M^2~.  \la{s1} \ee

The specific heat is defined as
\be
C_R = T \le(\f{\pa S_{BH}}{\pa T} \ri)_{R}~,
\ee
From (\ref{b1}) and (\ref{s1}), one obtains:
\be
C = 8\p M^2 \le( \f{R-2M}{3M-R} \ri)
= 4\p R^2 x^2 \le( \f{1-x}{3x-2} \ri)~.
\la{sp1}
\ee
From its form it is apparent that the specific heat is positive in the range
\be
\f{2}{3} \leq x \leq 1~~\Leftrightarrow~
2M \leq R \leq 3M~~,
\la{xrange}
\ee
which is possible only for the larger hole ($M_2$ above).  This can be
also observed from Figure 1.  For constant value of $R$, Figure 1 is a
relation between (inverse of) temperature and the masses of the two
Schwarzschild black holes. The mass of the larger solution increases
with increasing temperature. For the smaller hole it is just the
opposite. Also note that in the $R \rightarrow \infty$ limit, the
larger mass solutions do not exist meaningfully (it fills in the
cavity) irrespective of the temperature. Only the smaller solution
$x_1$ exists with $M=\frac{\beta}{8\pi}$ since $x_1\equiv
\frac{2M}{R}=\frac{\beta}{4\pi R}$ in this limit.
\begin{figure}[!h]
{\includegraphics[width=.42\textwidth]{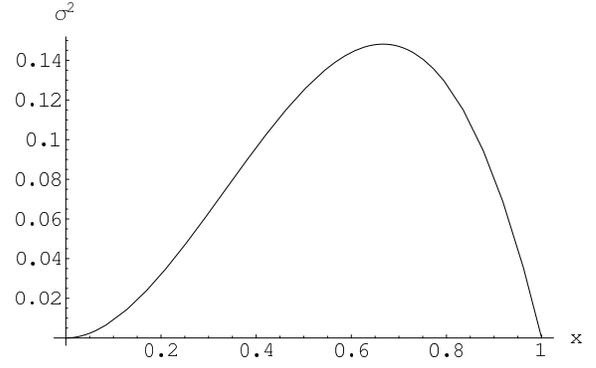}}
\\
\caption{For every value of $\sigma^2 \equiv 1/(4\pi
RT)^2<\frac{4}{{27}}$ there are two values of $x$, i.e., two black
holes. For the larger black hole $x\in [\frac{2}{3},1]$.}

\end{figure}
\subsection{\it Logarithmic Correction}
Substituting (\ref{sp1}) and (\ref{b1}) in (2), we obtain the
leading corrections to the entropy of the larger black hole:
\ba
&& \cals = S_{BH} -  \f{1}{2} \ln\left((3x-2)^{-1}\right)
+ \f{1}{2} \ln (4\pi) \cdots \nn  \\
      &=& S_{BH} -  \f{1}{2}
\ln\le[ \f{3}{R} \le( \f{S_{BH}}{4\p} \ri)^{1/2} -2 \ri]^{-1}
+\f{1}{2} \ln (4\p) + \cdots
\la{corr2}
\ea
Note that the first correction term is negative. This is because,
since $x \in [2/3,1]$, the argument of the logarithm is always greater
than unity.  Also, note that the corrections here is of a somewhat
different type than those found in the different applications of
(\ref{master0}). First, the correction depends on the value of $x$
which is a function of the product $RT$. When $RT \rightarrow
\frac{\sqrt{27}}{8\pi}$ the correction is enormous irrespective of the
value of $R$ alone (the horizon radius is $\frac{2R}{3}$). As one
increases the temperature, keeping the radius of the cavity fixed, the
infilling black hole becomes larger and fills in the whole cavity at
infinite temperature. However, the correction term decreases with
increasing temperature and still remains finite and non-zero
approaching $\ln (4\pi)$ at the infinite temperature limit. For small
cavity radius the correction term is therefore given by the first term
and is always appreciable and overpowers $S_{BH}$ if the temperature
is not high enough (signaling breakdown of formalism).  For large
cavity radius it is more appreciable at lower temperatures and becomes
negligible, albeit non-zero, at high temperature. Again, if $T$ is
near $\frac{\sqrt{27}}{8\pi R}$ it overpowers $S_{BH}$ even if the
latter is large (again signaling a breakdown).

Another way to see the different nature of the correction terms is to
note the coefficient of the correction term. The coefficient in front
of either of the two terms (which can be combined together) is $1/2$,
in contrast to BTZ of Schwarzschild anti-de Sitter black holes for
example, where the coefficients are $-3/2$ and $-d/2(d-2)$
respectively \cite{dmb}. However note that this $1/2$ is not in the
same footing as the coefficients in those cases as the correction term
here is not of the type $-k \ln(S_{BH})$. Rather, if we are to write
it in terms of $S_{BH}$ it takes the form given by the second equation
of (\ref{corr2}) which involves the cavity radius as well.

\subsection{\it Higher Dimensions}
Higher dimensional Schwarzschild black holes and their thermodynamics
within a finite isothermal cavity have been studied in \cite{mm}. For
$d$-dimensions the inverse temperature at the $(d-2)$-dimensional
boundary is given by:
\be
\b = T^{-1} = \f{4\p x R}{d-3} \sqrt{1 - x^{d-3}}
\equiv \f{4\p R\s}{d-3}~,
\la{scd}
\ee
from which one obtains:
\be
x^{d-1} - x^2 + \s^2 = 0~.
\ee
This equation cannot be solved using ordinary algebraic
methods. However, analytic solution is still possible \cite{mm}. This
would be needed if one requires the entropy correction as a function
of cavity radius and temperature. However, as in the above case of
four dimensions the general properties of the corrections to entropy
can be studied without requiring the explicit solutions by studying
how the correction term varies with $x$. (Again note that $x$ takes
value within the interval $[0,1]$.)  This will be done in the next
section as a special case of $d$-dimensional Reissner-Nordstr\"om. As
we will see there entropy corrections to $d$-dimensional
Reissner-Nordstr\"om (and hence Schwarzschild) black holes can be
exactly evaluated.

\section{Reissner-Nordstr\"om Black Holes}
\la{secrn}
We now extend the above analysis for the Reissner-Nordstr\"om
black holes which have a charge. This means we would have to specify
suitable electrostatic potential at the boundary. We first discuss
the case of four dimensions.
\subsection{\it Four dimensions}
The Euclideanised  Reissner-Nordstr\"om black hole metric is:
\ba
ds^2 &=&  \le( 1 - \f{2M}{r} + \f{Q^2}{r^2} \ri) d\t^2 \nn \\
&+&  \le( 1 - \f{2M}{r} + \f{Q^2}{r^2} \ri)^{-1} dr^2 + r^2 d\O_2^2~,
\ea
with inverse temperature at boundary \cite{york2}:
\ba
\b = T^{-1} &=& 4\p r_+ \le( 1 - \f{Q^2}{r_+^2} \ri)^{-1}
\le(1 - \f{r_+}{R} \ri)^{1/2} \le(1 - \f{Q^2}{r_+ R} \ri)^{1/2} \nn \\
 &=& \f{4\p R x^{5/2}}{x^2-\a^2}  \sqrt{(1-x)(x-\a^2)} ~,
\ea
where $r_+ = M + \sqrt{M^2-Q^2}$ is the radius of the outer horizon and
$$ \a = \f{Q}{R} ~.$$
The counterpart of Eq.(\ref{cube}) is:
\be
(1-\Phi^2)x^3 - x^2
+ (1-\Phi^2)^2 \s^2 = 0 ~,
\la{equil1}
\ee
where $x$ and $\s$ has identical definitions in terms of $r_+$ and
$\b$ as before.
%
%
Here, $\Phi$ is the potential difference between $r_+$ and $R$,
suitably red-shifted:
\ba
\Phi &=& \f{Q}{r_+} \le(1- \f{r_+}{R} \ri)^{1/2}
\le( 1 - \f{Q^2}{Rr_+} \ri)^{-1/2} \nn \\
&=& \a~\sqrt{\f{1-x}{x^2 - \a^2 x}}~,
\ea
Now, the two positive solutions of $x$ are:
\ba
x_1 &=& \f{1}{3(1-\Phi^2)}
\le( 1 - 2 \cos \le( \f{\a}{3}  + \f{\p}{3} \ri) \ri)  \\
x_2 &=& \f{1}{3(1-\Phi^2)} \le( 1 + 2 \cos\f{\a}{3} \ri)  \\
\cos\a &=& 1 - \f{27}{2} \s^2 (1-\Phi)^4 ~,
\ea
of which only the second one is thermodynamically stable. The entropy is,
in this case:
\be
S_{BH} = \p r_+^2 = \p R^2 x^2 ~.
\ee
(where we have dropped the subscript $2$ from $x_2$).
The specific heat at constant $\Phi$ and $R$ can now be computed
by using the relation:
\be
C_{\Phi,R} = \le(\f{\pa E}{\pa T }\ri)_{\Phi,R}
= - \le( \f{\pa S/\pa x}{\pa \ln\s/\pa x} \ri)_{\Phi,R}
\la{easy1}
\ee
with $\s$ obtained from Eq.(\ref{equil1}). This yields:
\be
C_{\Phi,R} = \f{4\p R^2 x^3 (1-x)}{3x^2 - 2x - \a^2}~,
\ee
where now the specific heat is non-negative in the range:
\be
r_+ \leq R \leq  \f{3}{2}r_+ - \f{1}{2} \f{Q^2}{r_+}~,
\ee
Equivalently:
\ba
x &\leq& 1 ~,\\
3x^2 - 2x - \a^2 &\geq& 0 ~.
\ea
In the above range, the corrected entropy assumes the form:
\ba
{\cal S} &=& S_{BH} -
\f{1}{2} \ln\le( \f{x^2 (3x^2-2x-\a^2)(x-\a^2)}
{(x^2-\a^2)^2} \ri)^{-1} \nn \\
&+& \f{1}{2} \ln (4\p) + \cdots
\ea
This can be expressed in terms of the Bekenstein-Hawking entropy as:
\ba
&&{\cal S} = S_{BH} \nn \\
&-& \f{1}{2}
\ln\le( \f{
\le( \f{3S_{BH}}{\p R^2}  - 2\sqrt{\f{S_{BH}}{\p R^2}} - \a^2 \ri)
\le( \sqrt{\f{S_{BH}}{\p R^2}} - \a^2 \ri) }
{
\le( \f{\p R^2}{S_{BH}} \ri)
\le(  \f{S_{BH}}{\p R^2}  -\a^2  \ri)^2 }  \ri)^{-1} \nn\\
&+& \f{1}{2} \ln (4\p) + \cdots
\ea
As expected, the above reduces to Eq.(\ref{corr2}) for $\a = 0$.

\subsection{\it Higher dimensions}
For the $d$-dimensional Reissner-Nordstr\"om black hole the Euclidean metric
is given by
\ba
&ds&^2 =
\le(  1 - \f{16\p M}{(d-2) \O_{d-2} r^{d-3}}
+ \f{16\p Q^2}{(d-2)(d-3) r^{2(d-3)}} \ri) d\t^2 \nn \\
&+& \le(  1 - \f{16\p M}{(d-2) \O_{d-2} r^{d-3}}
+ \f{16\p Q^2}{(d-2)(d-3) r^{2(d-3)}} \ri)^{-1} dr^2  \nn \\
&& + r^2 d\O_{d-2}^2 ~,
\ea
where $d\O_{d-2}^2$ is a unit $S^{d-2}$ sphere with the standard round
metric on it. The inverse temperature is given by:
\ba
\b &=& T^{-1} \nn \\
&=& \f{4\p r_+}{d-3}  \le( 1 - \f{2Q^2}{(d-2)(d-3) r_+^{2(d-3)}} \ri)^{-1}
~\times \nn\\
&& \le( 1 - \le(\f{r_+}{R}   \ri)^{d-3} \ri)^{\f{1}{2}}
\le(1  - \f{2Q^2}{(d-2)(d-3) (r_+ R)^{d-3}}  \ri)^{\f{1}{2}} \\
&=& \f{4\p R x^{1 + 3(d-3)/2}}{d-3} \le(  x^{2(d-3)} - \a^{2(d-3)}\ri)^{-1}
~\times \nn \\
&{}& \le( 1 - x^{d-3} \ri)^{1/2} \le( x^{d-3} - \a^{2(d-3)}\ri)^{1/2}~,
\la{rndtemp}
\ea
where $x=r_+/R$ as before, and now $\a$ is defined as:
$$ \le(\a R \ri)^{2(d-3)} = \f{2Q^2}{(d-2)(d-3)}~. $$
Similarly, the the potential difference is:
\ba
\Phi &=& \f{Q}{r_+^{d-3}}
\sqrt{\f{1 - (r_+/R)^{d-3}}{ 1 - 2Q^2/(d-2)(d-3)(r_+R)^{d-3}} } \\
&=& \sqrt{\f{2}{(d-2)(d-3)}}~{\a^{d-3}}
\sqrt{\f{1-x^{d-3}}{x^{2(d-3)} - \a^{2(d-3)} x^{d-3}}}
\ea
and Eq.(\ref{equil1}) is generalised to:
\ba
&& \le( 1- \f{(d-2)(d-3)}{2} \Phi^2 \ri)x^{d-1} - x^2  \nn \\
&+& \le(1-\f{(d-2)(d-3)}{2} \Phi^2 \ri)^2 \s^2 = 0 ~,
\la{equil2}
\ea
where $\s$ has been defined as in Eq.(\ref{scd}).
Using (\ref{easy1}), the specific heat at constant
$R$ and $\Phi$ can now be computed as:
\be
C_{\Phi,R} = \f{(d-2) \O_{d-2} R^{d-2}~x^{2d-5} \le( 1 - x^{d-3} \ri)}
{(d-1) x^{2(d-3)} - 2x^{d-3} - (d-3) \a^{2(d-3)}}~.
\la{rndc}
\ee
Stability is now guaranteed if the parameters lie in a range such that:
\be
(d-1) x^{2(d-3)} - 2x^{d-3} - (d-3) \a^{2(d-3)} \geq 0 ~.
\ee
Explicit evaluation of $x(M,Q)$ and $R (M,Q)$ from the above
is complicated for
higher dimensions though, because of the higher powers of $x$.
Eq.(\ref{rndc}), along with (\ref{rndtemp}) now
gives the corrected entropy as:
\ba
\cals &=& S_{BH} \nn \\
&-& \f{1}{2}
\ln [~x^{d-2} \le( (d-1) x^{2(d-3)} - 2x^{d-3} - (d-3) \a^{2(d-3)} \ri)
\nn \\
&& \times~\f{ x^{d-3} - \a^{2(d-3)} }
{ R^{d-4} \le( x^{2(d-3)} - \a^{2(d-3)} \ri)^2 }~]^{-1}  \nn \\
&+& \f{1}{2} \ln \le[\f{(d-2)\O_{d-2}}{(4\p)^2} \ri]+ \cdots
\ea
which can be expressed in terms of $S_{BH}$ as:
\ba
&& \cals = S_{BH} \nn \\
&-& \ln [ (\sbh) ~\times \nn \\
&& \{ (d-1) (\sbh)^{2(d-3)/(d-2)} \nn \\
&& - 2 (\sbh)^{(d-3)/(d-2)}
- (d-3) \a^{2(d-3)}  \} \times  \nn \\
&& \f{\{ (\sbh)^{(d-3)/(d-2)} - \a^{2(d-3)}  \}}
{\{ (\sbh)^{2(d-3)/(d-2)} - \a^{2(d-3)} \}^2} ~]^{-1} \nn \\
&+& \f{1}{2} \ln \le[\f{(d-2)\O_{d-2}}{(4\p)^2} \ri]+ \cdots
\ea
For zero-charge, that is for $d$-dimensional Schwarzschild
black holes, the corresponding expression is:
\ba
\cals &=& S_{BH}
- \f{1}{2} \ln\le[
\f{(d-1)(\sbh)^{(d-3)/(d-2) -2 }  }{R^{d-4} (\sbh)^{(d-4)/(d-2)} }
\ri]^{-1} \nn \\
&+& \f{1}{2} \ln \le[\f{(d-2)\O_{d-2}}{(4\p)^2} \ri]+ \cdots
\ea
In each case, the entropy correction is different in nature than 
in previously studied examples.

\section{Discussions}
\la{secdis}

In this article, we have shown how leading order corrections due to
thermal fluctuations can be obtained for Schwarzschild and
Reissner-Nordstr\"om black holes within a finite cavity. With
appropriate boundary data, within a cavity thermodynamically stable
solutions are possible. Normally, such corrections cannot be computed
for Schwarzschild owing to its negative specific heat.
For Reissner-Nordstr\"om black holes, corrections of this type were
computed in \cite{dmb} for a very small range of parameters near
extremality, and that too in a perturbative approach (with
perturbation parameter $Q/M$). Here, the result is non-perturbative,
and holds for any parameter range.
Finally, we generalised the stability analysis to
higher dimensional Reissner-Nordstr\"om black holes, and computed its
leading order entropy corrections.  
We found that the leading logarithmic corrections come with a negative
signature.
This means
that the corrected dimensionality of the Hilbert space of the black
holes under consideration, which is ${\cal N} = \exp(\cals)$, is smaller
than the uncorrected dimension.  It would be interesting to
explore implications of this result.  It would also be interesting to
do the generalisation to other types of black holes, e.g. Kerr-Newmann
as well as black holes which are not asymptotically flat, e.g.  those
that are asymptotically anti-de Sitter or de-Sitter. The implications
of the latter to $AdS/CFT$ and the $dS/CFT$ correspondence may turn
out to be interesting. We hope to report on this elsewhere.

\vs{.4cm} \no {\bf Acknowledgements} SD thanks R. K. Bhaduri and
P. Majumdar for useful correspondence and interesting discussions.
This work is supported in part by the Natural Sciences and Engineering
Research Council of Canada. MMA would like to thank the Overseas
Research Scheme, the Cambridge Commonwealth Trust and DAMTP for
financial support.

\end{multicols}

\end{document}